\documentclass[a4paper,11pt]{article}
\usepackage{pos}
\usepackage{braket}
\usepackage{svg}
\usepackage{pgf}
\usepackage{graphicx} 
\usepackage[printwatermark]{xwatermark}
\newwatermark[firstpage,color=black!50,angle=270,scale=0.5,xpos=120,ypos=90]{MS-TP-21-32}

\title{$B_s \to D^{(*)}_s$ form factors from lattice QCD with ${ N_f}=2$ Wilson-clover quarks}

\author[a]{Beno\^it~Blossier}
\author[b]{Pierre-Henri~Cahue}
\author[c]{Jochen~Heitger} 
\author[b]{Simone~La~Cesa}
\author*[a,c]{Jan Neuendorf}
\author[d]{Savvas Zafeiropoulos}

\affiliation[a]{Laboratoire de Physique des 2 Infinis Ir\`ene Joliot-Curie, CNRS/IN2P3, Universit\'e Paris-Saclay,\\  B\^atiment 210, 91405 Orsay Cedex, France}
\affiliation[b]{Laboratoire de Physique Subatomique et de Cosmologie,\\ CNRS/IN2P3, 4 All\'ee des Martyrs, 38000 Grenoble, France}
\affiliation[c]{Institut~f\"ur~Theoretische~Physik, Westfälische Wilhelms-Universität Münster,\\ Wilhelm-Klemm-Str.~9, 48149~M\"unster, Germany}
\affiliation[d]{Aix Marseille Univ, Universit\'e de Toulon,\\ CNRS, CPT, Marseille, France}

\emailAdd{blossier@ijclab.in2p3.fr}
\emailAdd{cahue@lpsc.in2p3.fr}
\emailAdd{heitger@wwu.de}
\emailAdd{simo.lcs@gmail.com}
\emailAdd{jan.neuendorf@wwu.de}
\emailAdd{savvas.zafeiropoulos@cpt.univ-mrs.fr}

\abstract{
We report on a two-flavour lattice QCD determination of the $B_s\to D_s$
and $B_s\to D_s^*$ transitions, which in the heavy quark limit can be
parameterised by the form factors $\mathcal{G}$, and $h_{\text A_1}$, $h_{\text A_2}$ and
$h_{\text A_3}$. In the search of New Physics through tests of lepton-flavour
universality, $B_s$ decay channels are complementary to $B$ decays and
widely studied at $B$ factories and LHCb. The purpose of our study is to
explore a suitable method to extract form factors associated with $b\to c$
currents from lattice QCD. In particular, we present numerical results for
$\mathcal{G}$ and $h_{\text A_1}$.

}

\FullConference{%
 The 38th International Symposium on Lattice Field Theory, LATTICE2021
  26th-30th July, 2021
  Zoom/Gather@Massachusetts Institute of Technology
}

\newcommand{\Bs}{{B_s}}
\newcommand{\Ds}{{D_s}}
\newcommand{\ha}{{h_{A_1}}}

\begin{document}
\maketitle

\section{Motivation and preliminary remarks}
The purpose of this work is to study whether Wilson-Clover fermions, 
in combination with the step-scaling in mass method \cite{Blossier:2009hg, Atoui:2013mqa}, allow for the extraction of reliable results for $B$ decays,
as far as cut-off effects and contamination by excited 
states are concerned.
We perform an analysis on $N_f=2$ ensembles, created by the CLS effort.
The work presented here is described in more detail in the corresponding article \cite{blossier2021extraction}.

We consider the semi leptonic decay $B_s$ to $D_s^{(*)}$, as sketched in fig. \ref{schematic}. 

\begin{center}
    \vspace{40pt}
    \includegraphics[width=0.7\textwidth]{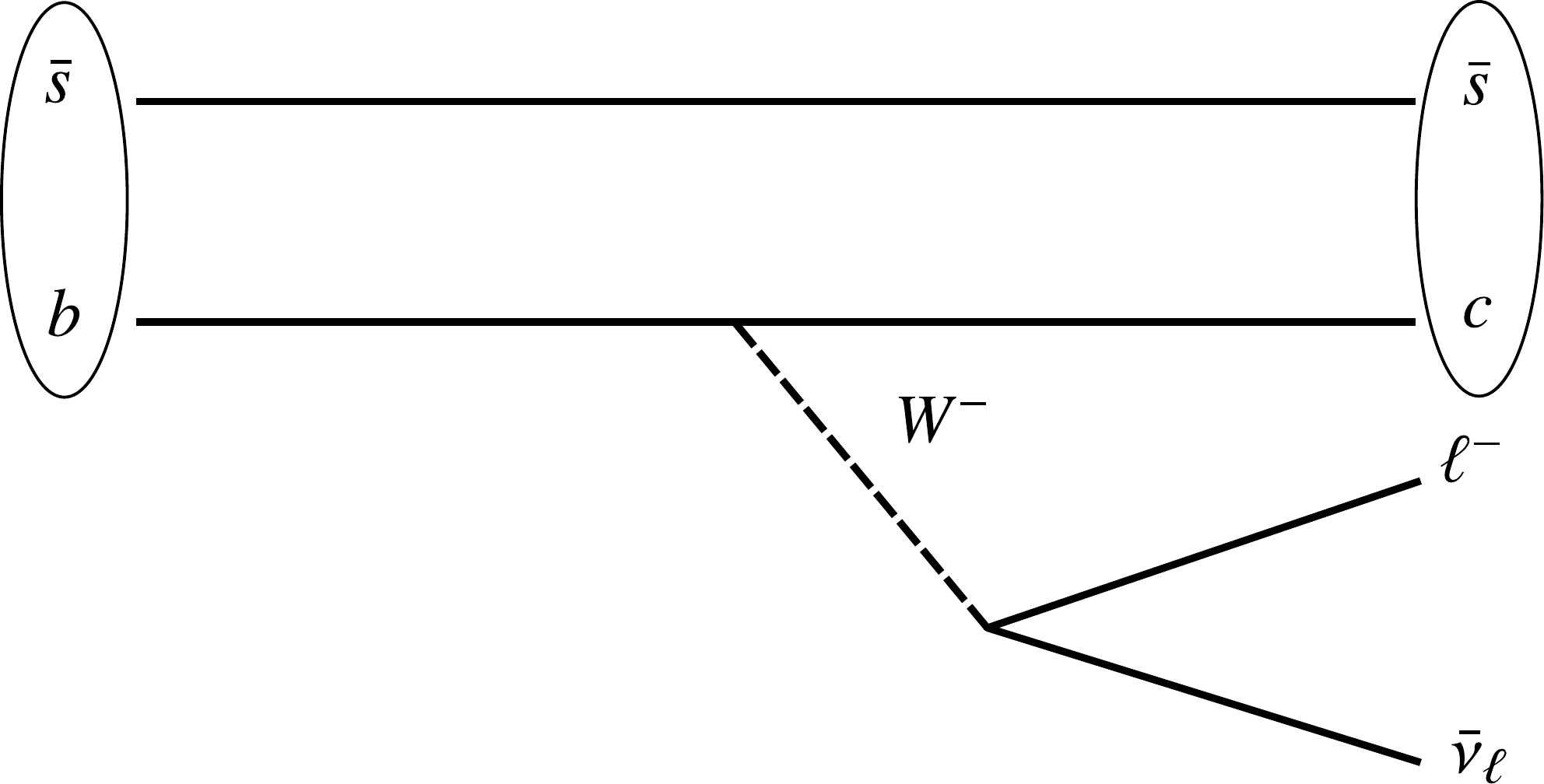}
  
    \captionof{figure}{\label{SM_semilep-diagram}
        Schematic of the decay $\bar{B^0_s}\to D_s^{+(*)} \ell^- \bar{\nu}_{\ell}$
        \label{schematic}}
    \vspace{40pt}
\end{center}
The decay width includes the CKM matrix element $V_{cb}$ as well as a form factor that encodes the long-distance dynamics of QCD. 
Our main focus is on $\Bs\to\Ds$, where this factor is called $\mathcal{G}$.
\begin{equation}
    \frac{d\Gamma_{\Bs\to \Ds}}{d w}\propto|V_{cb}|^2\cdot |{\mathcal{G}(w)} |^2 
    \transparent{1}\cdot {G}^2_F\transparent{1}(m_{\Bs}\transparent{1}
    +m_\Ds\transparent{1})^2\transparent{0.1}
    \transparent{1},
\end{equation}
where $w=\frac{E_\Ds}{m_\Ds}$ is the relative velocity of the $\Ds^{(*)}$ meson. It is $h_{\text A _1}$ for $\Bs\to\Ds^*$.

The strong contribution only depends on $w$. 
It is suitable to describe physics of heavy-light mesons by means of Heavy Quark Effective Theory (HQET). In first approximation, the light degrees of freedom
live in the potential created by a static source of colour.
We are interested in the zero recoil case, where $w=1$. In the heavy quark limit, Heavy
Quark Symmetry implies that $\mathcal{G}^{m_h \to \infty}(1)=1$.
Since $b$ and $c$ are heavy, we can predict ab initio that $\mathcal{G}(1)$ is close to one.

The decay can be parametrised as  
\begin{equation}
    \braket{D_s(k)|\bar{b}\gamma_\mu c|B_s(p)}
    =A_\mu(p,k){\mathcal{G}(w)}+B_\mu(p,k)f_0(w),\label{formfac}
\end{equation}
where all dependencies other than the momentum transfer have been absorbed.
The left hand side is a matrix element, which can be accessed on the 
lattice \cite{Atoui:2013mqa}. 

\section{Mass step-scaling}
To control cut-off effects we have decided not to compute
our observables directly at the b-quark scale.
Instead, on each ensemble, a set of six meson masses $m_{h_{i\in [0,..,5]}s}$ is chosen
to fulfill,
\begin{align}
    &m_{h_0 s}=m_{D_s^{}},\\ 
    &\frac{m_{h_{i+1} s}}{m_{h_i s}}=\lambda\quad \text{with} \quad \lambda=\sqrt[6]{\frac{m_\Bs}{m_\Ds}}.
\end{align}
The bare heavy quark mass is tuned to ensure these relations. 

This approach has more advantages than just allowing us to extrapolate our results to 
the $B_s$. As mentioned in the previous section, 
$\mathcal{G}(w=1)$ equals one at the elastic point $m_{h_0 s}=m_{D_s}$.
We exploit this, by expressing $\mathcal{G}^{\Bs\to\Ds}(w=1)$ as:
\begin{align}
    \mathcal{G}^{\Bs\to\Ds}(w=1)&=\mathcal{G}_{i=6}(1)=\frac{\mathcal{G}_6(1)}{\mathcal{G}_5(1)}
    \cdot\frac{\mathcal{G}_5(1)}{\mathcal{G}_4(1)}\cdot...\cdot \frac{\mathcal{G}_1(1)}{\mathcal{G}_0(1)}\cdot \mathcal{G}_0(1)\\ 
    &=\sigma_6\cdot\sigma_5\cdot ... \cdot \sigma_1 \quad\text{with}\quad \sigma_i=\frac{\mathcal{G}_i(1)}{\mathcal{G}_{i-1}(1)}
\end{align}
We only need to consider the ratios of $\mathcal{G}$'s, which leads to the cancellation 
of correlated errors and renormalization constants. But since we do not 
compute $\mathcal{G}_6$, we do not get a value for $\sigma_6$ either. 
\begin{figure}
    \begin{center}
        \includegraphics[width=0.48\textwidth]{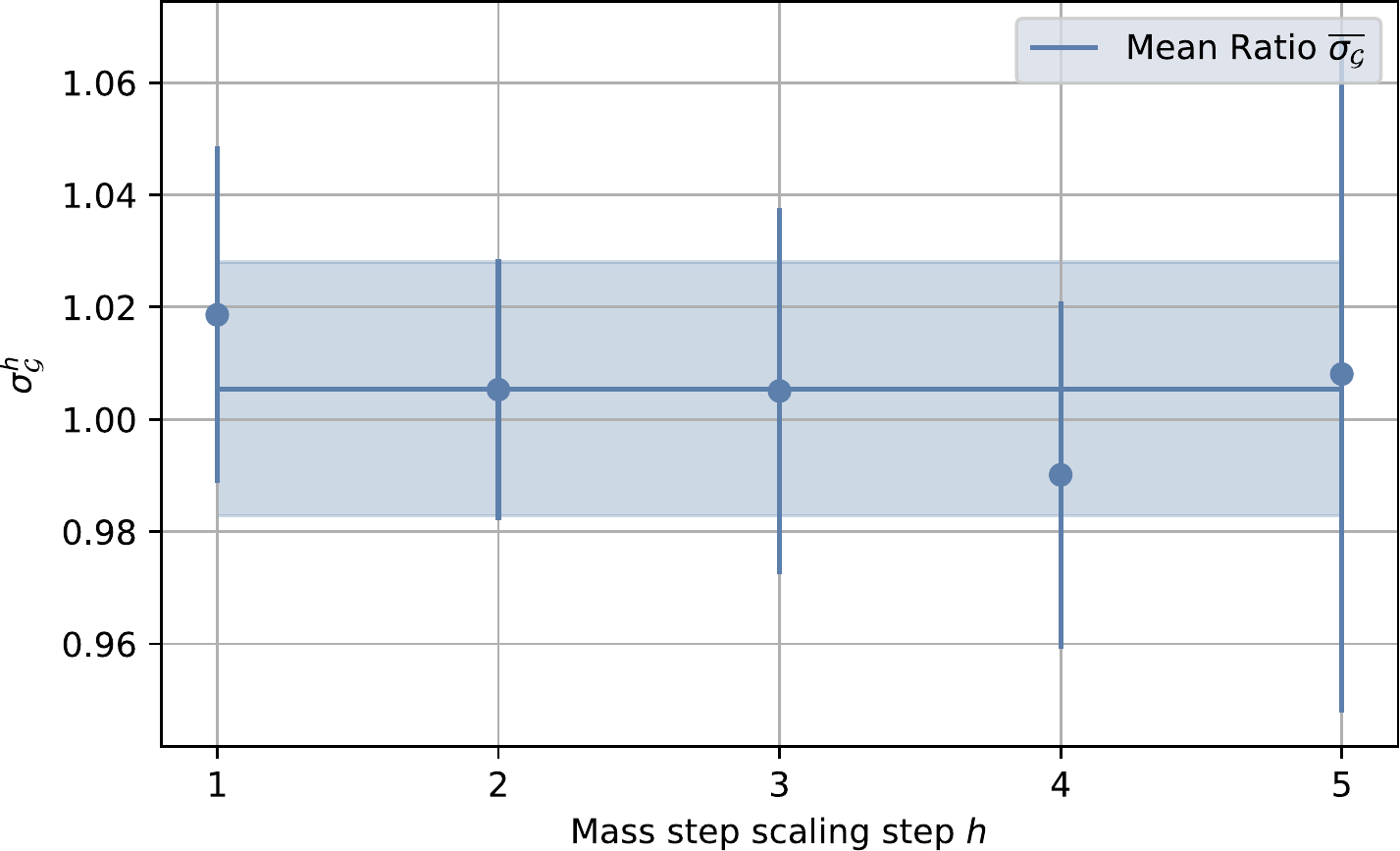}
        \caption{\label{fig_ratiofit} Ratios $\sigma$ at the different mass step-scaling steps.}
    \end{center}
\end{figure}
This means that we have to perform some extrapolation to gain access to it. 
In Fig.\,\ref{fig_ratiofit} we show that our ratios are compatible with a constant. 
Therefore, for our final results in section \ref{resultsection}, we opted for computing 
our final result from the mean value $\bar{\sigma}$ as $\mathcal{G}^{\Bs\to\Ds}(w=1)=\overline{\sigma}^6$.

For the computation of $\ha$ we can gain no such advantage, since 
$\ha^{\Ds^*\to\Ds}\neq 1$. Nevertheless, we use the same approach 
for consistency. 

\section{Matrix elements}
Looking at the left hand side of eq.\,(\ref{formfac}), we can see the form 
$\braket{P|V|P}$. On the lattice, we compute a corresponding three-point correlation 
function:
\begin{align}
\nonumber
C^{\theta}_{ij}({\textstyle \frac{T}{2}} ],t) &=\sum_{\vec{x},\vec{y}} \langle P_{hs,\, i}({\textstyle \frac{T}{2}}) V^{I,\, \theta}_{hc,\, \mu}(t)P^{\dag\, \theta}_{cs,\,j}(0)\rangle\\
    &\stackrel{1\ll t\ll {\textstyle \frac{T}{2}}}{\sim} \frac{Z_{H_s} Z_\Ds}{4 E_{H_s}E_{\Ds^{}}}e^{-E_{H_s}t}e^{-E_{\Ds^{}}(\frac{T}{2}-t)}\braket{D_s(k)|\bar{b}\gamma_\mu c|H_s(p)},
\label{3pt}
\end{align}
where $Z$ is the source amplitude.
The bare matrix element at the left hand side of  eq.\,(\ref{formfac}) appears.
The source-sink-separation is kept constant at $\frac{T}{2}$, which is $\gtrsim 2\,{\rm fm}$ for the studied ensembles.

Using the two point correlators belonging to the propagation of $H_s$ and ${\Ds^{(*)}}$, 
we can extract the energies and amplitudes and eliminate them from eq.\,(\ref{3pt}).

To optimise for a better overlap with the respective ground states of $\Ds^{(*)}$ and $H_s$, 
we made use of multiple smeared sources and sinks and the generalised eigenvalue problem (GEVP).
The (ground state) eigenvectors were computed for the two point correlators. 
They could then also be used to project the three-point correlator (eq.\,(\ref{3pt})), 
since the setup of the sources is consistent.

\section{Lattice details}
Our data were computed on eight different CLS ensembles with $N_f=2$ 
and $O(a)$ improved Wilson-Clover fermions \cite{Sheikholeslami:1985ij,Luscher:1996ug}.
\begin{center}
    \vspace{10pt}
    \includegraphics[width=0.6\textwidth]{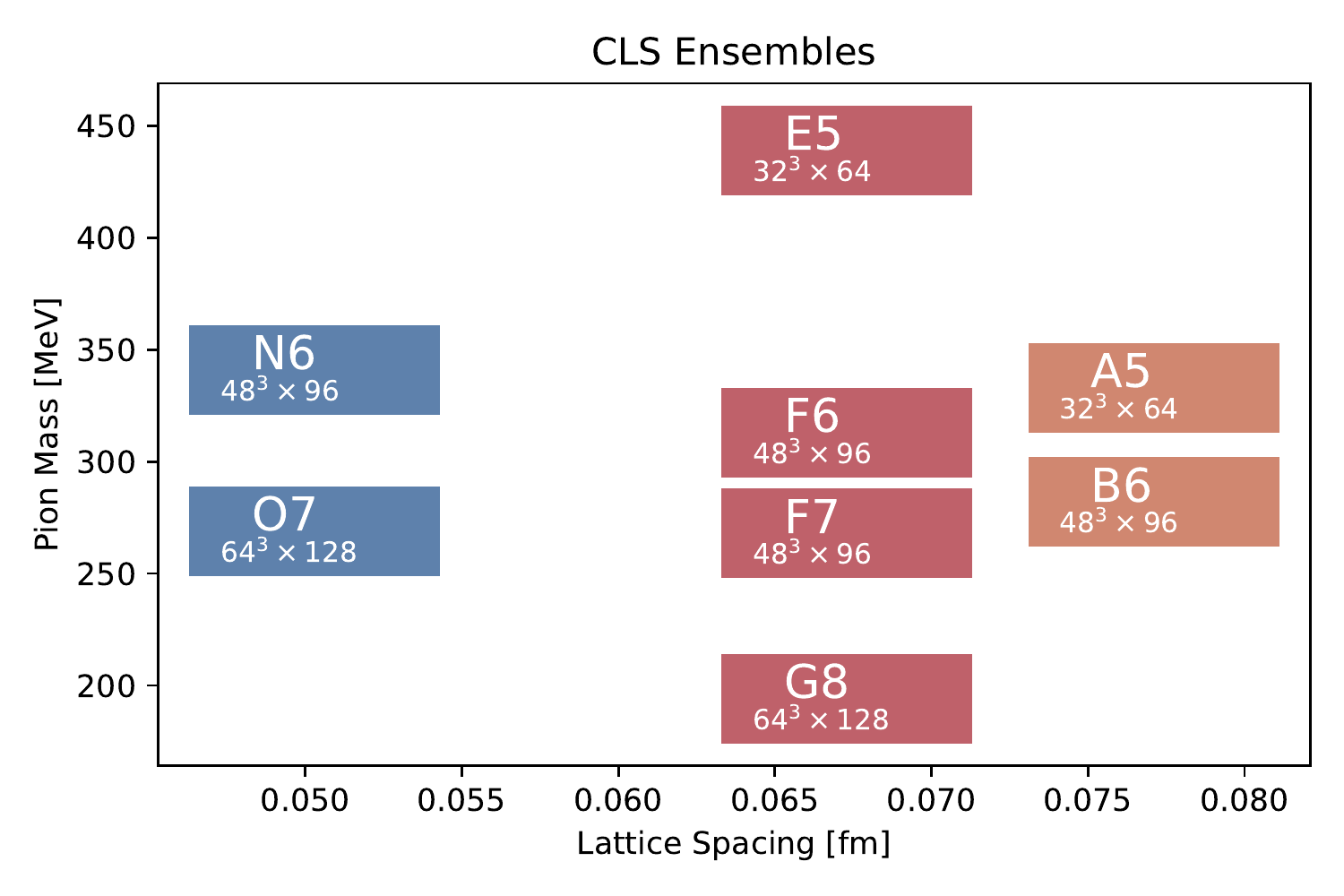}
    \captionof{figure}{Parameters of the used CLS ensembles.\label{CLS}}
\end{center} 
The parameters of these ensembles are shown in Fig.\,\ref{CLS}. 
Hopping-parameters corresponding to $c$- and $s$-quarks, 
as well as the appropriately tuned values for the mass step-scaling 
steps are known from previous works \cite{Fritzsch_2012,heitger2013charm}.

We are interested in the case of zero recoil ($w=1$). 
The kinematic vanishing of the spatial matrix element 
$\langle\Ds|V^i|\Bs\rangle$ makes it 
impossible to extract 
$\mathcal{G}$ at $w=1$ directly. For $\ha$ we do not have this problem. 
We need to calculate $\mathcal{G}$ with different momentum transfers and extrapolate to zero recoil.
To introduce momenta to the $\Ds$, we impose spatial isotropic twisted-boundary conditions
with a twisting angle $\theta$ to the $c$-quark. 
On each ensemble we 
compute the correlators for six values of $\theta$ as well as without twisting. 
\begin{figure}
    \centering
    \includegraphics[width=0.47\textwidth]{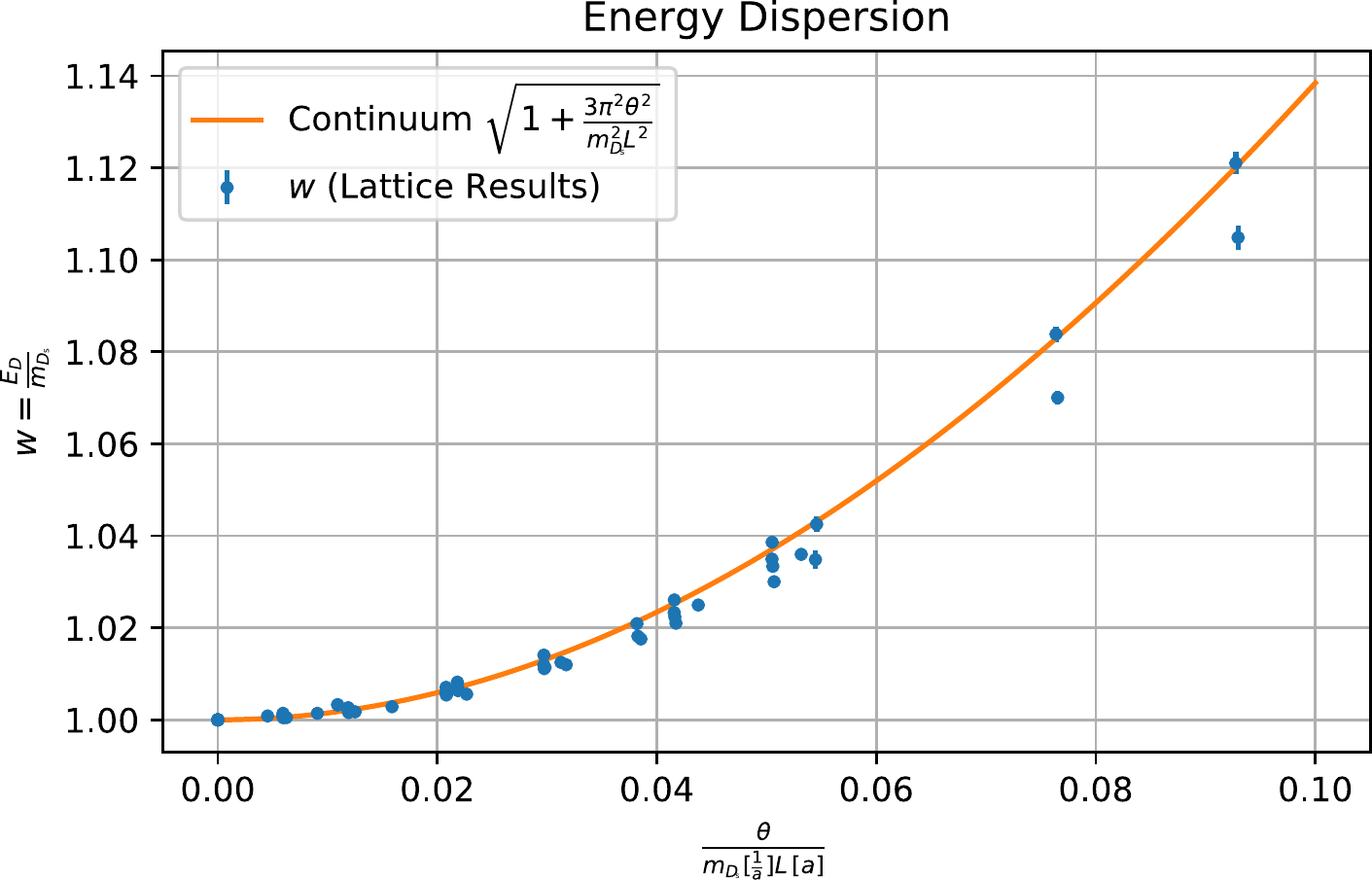}\hspace{0.05\textwidth}
    \includegraphics*[width=0.47\textwidth]{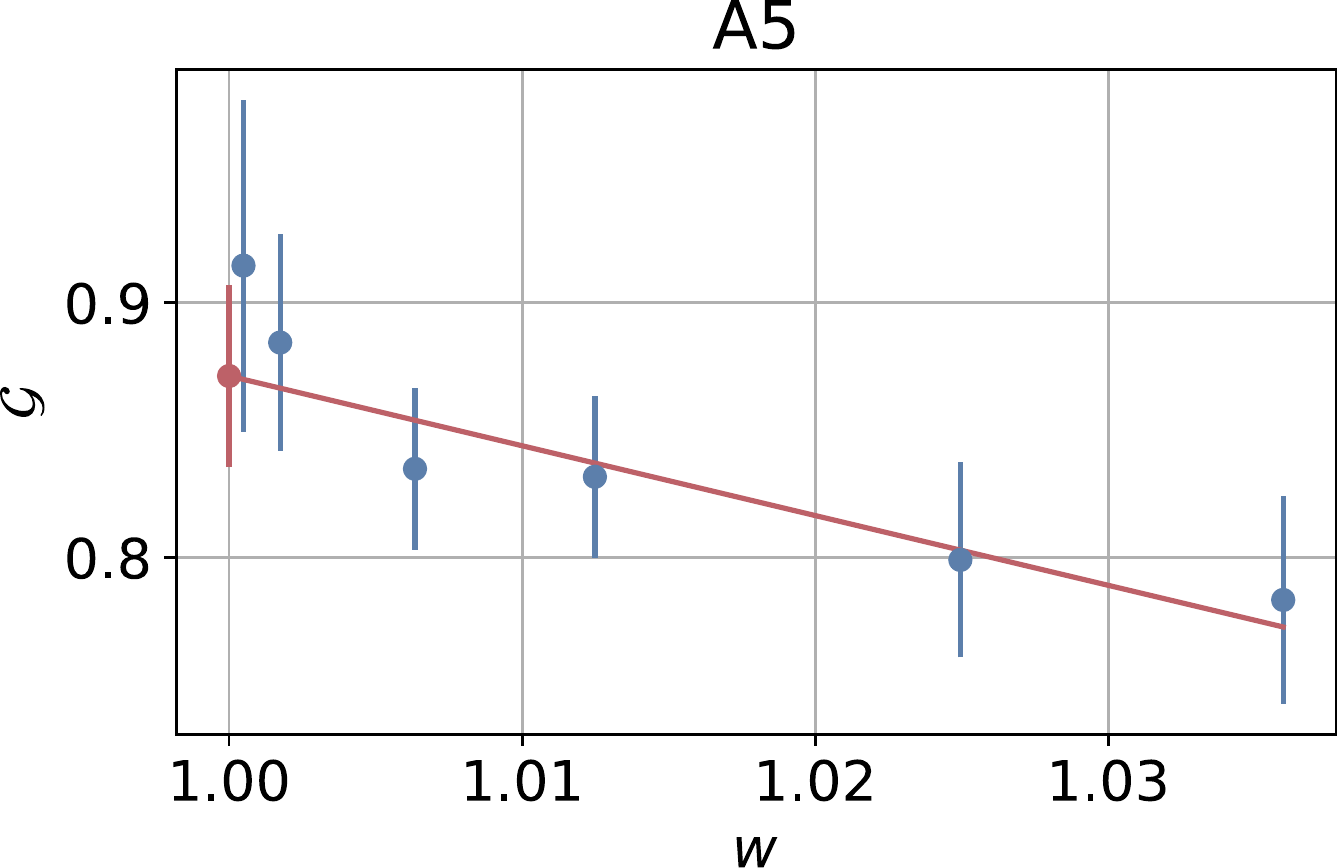}
    \caption{(\textit{left}) \label{disp} Dispersion relation for $D_s$ with twisted boundary conditions. (\textit{right}) \label{wto1} Exemplary extrapolation of $\mathcal{G}$ to $w=1$.}
\end{figure}
In Fig.\,\ref{disp} we see that the introduction of a spatial twisting-angle for the $c$-quark does indeed 
inject the $\Bs$ with the expected momentum. Here $w=\frac{E_\Ds}{m_\Ds}$ is computed directly from the spectroscopy 
of the two point correlation functions. 
We obtain $\mathcal{G}$ for all values of $\theta$ and perform a linear extrapolation in $(w-1)$. 
An example for this is shown in Fig.\,\ref{wto1}.

\section{Results at the physical point\label{resultsection}}
For all observables $o$, which need to be extrapolated 
in this section, we use the following ansatz: 
\begin{align}
        &o(a,m^2_\pi)=o_0 + o_1 \times (m^2_\pi / m^{2, {\rm physical}}_\pi)+o_2 \times (a/a_{\beta=5.3})^2 \label{elastic_fitfunc}\\
        \implies &o^{\rm physical}=o(0,m^{2, {\rm physical}}_\pi)=o_0+o_1
\end{align}
For some quantities, other parameters such as heavy mass dependency and a dependency on a possible mistuning of the mass step-scaling steps has been tried as well \cite{blossier2021extraction}. 

We have asserted that $\mathcal{G}(1)$ is equal to one at the elastic point. 
This can be shown from the lattice data by performing an extrapolation 
to the physical point. 
\begin{figure}
    \begin{center}
        \includegraphics[width=0.48\textwidth]{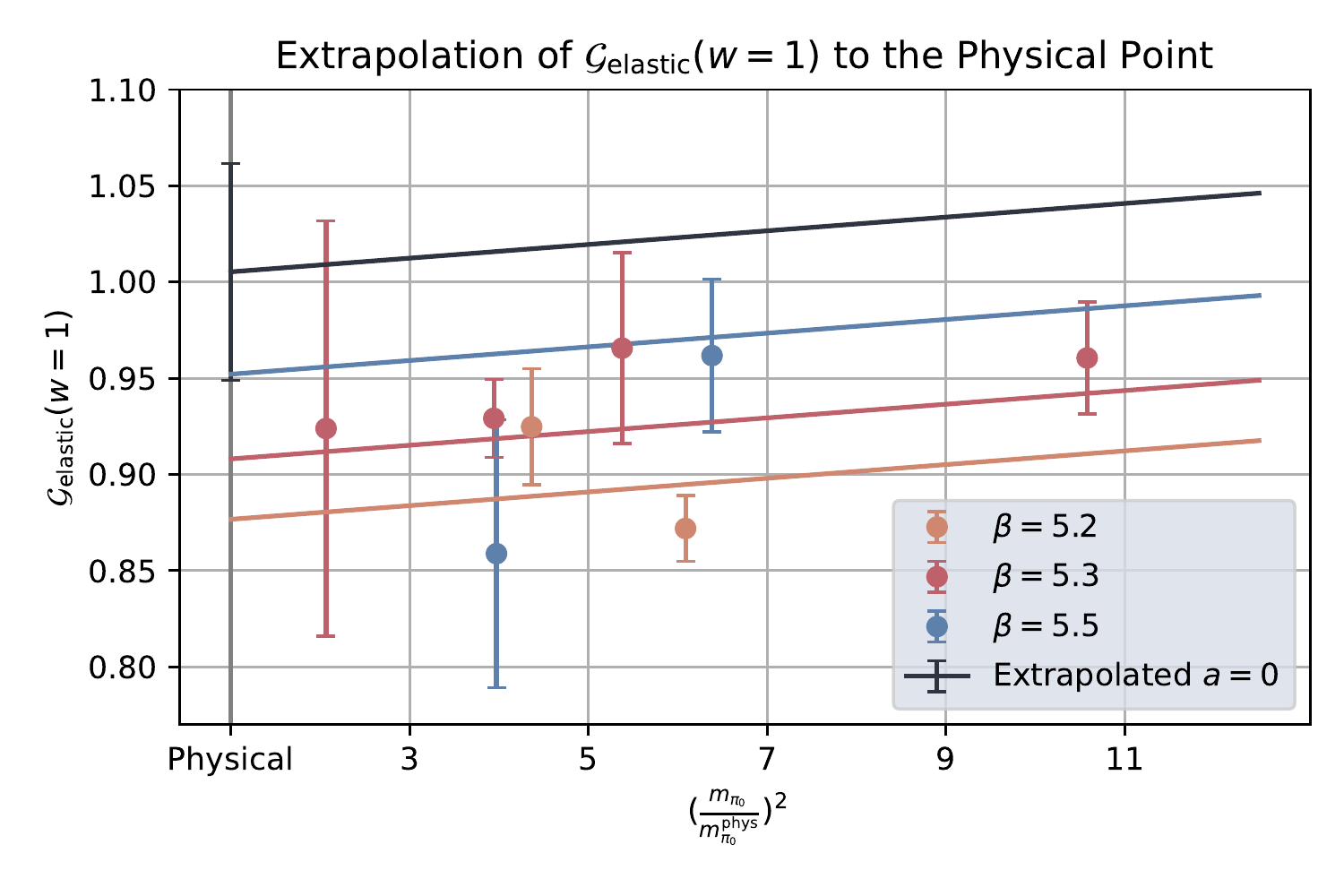}
        \caption{\label{fig_elastic} Visualisation of the combined fit eq.\,(\ref{elastic_fitfunc}) and extrapolation to the physical point for ${\mathcal{G}}_{\mathrm{elastic}}$.  }
    \end{center}
\end{figure}
The result of this extrapolation is shown in Fig.\,\ref{fig_elastic}. 
It is indeed compatible with one. 

In the case of $\mathcal{G}$ we only need to consider the ratios between successive
mass step-scaling steps. These are computed directly 
on every ensemble to ensure the cancellation of correlated errors. 
Then the ratios are extrapolated (separately for every mass step-scaling step)
to the physical point according to eq.\,(\ref{elastic_fitfunc}). 
A combined fit with additional parameters has been tried as well with compatible results. 
The resulting ratios at the physical point are shown in Fig.\,\ref{fig_ratiofit}.
\\We get a final result for $\mathcal{G}$:
\begin{equation}
    \label{final_result_G}
     \mathcal{G}^{B_s \to D_s}(w=1)=\prod_{i=1}^6 \sigma_i \approx\overline{\sigma}^{6}=1.03(14)
\end{equation}
For $\ha$ we obtain
\begin{align}
    h^{B_s \to D^*_s}_{A_1}(1)&=h^{D_s \to D^*_s}_{A_1}(1) \times \prod_{i=1}^6 \sigma_i^{\ha} \approx h^{D_s \to D^*_s}_{A_1}(1) \times (\overline{\sigma^{{h}_{A_1}}} )^6\\
    &=0.825(83)\times (1.005(23))^6=0.85(16).
\end{align}

\section{Discussion and conclusion}
In this exploratory $N_f=2$ study, we have obtained the form factors $\mathcal{G}(w=1)=1.03(14)$ and 
$h_{A_1}(w=1)=0.86(16)$, associated with the semileptonic decays $\Bs\to\Ds$ and $\Bs\to\Ds^*$ 
respectively, using the method of step-scaling in mass.

\begin{figure}
	\centering
	\resizebox{0.95\textwidth}{!}{\input{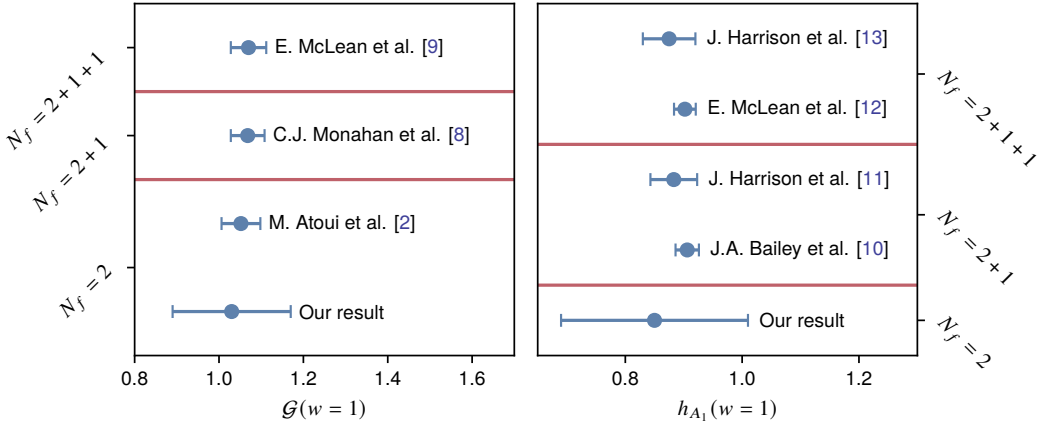}}
	\caption{Comparison to previous results for $\mathcal{G}$ (\textit{left}) and $h_{{\rm A}_1}$ (\textit{right}). \label{fig_comparison}}
\end{figure}
From Fig.\,\ref{fig_comparison}, where we show a comparison to previous results, one infers that we are facing substantial statistical uncertainty. 
This might in part be explained by the large source-sink-separation. 
Moreover, the mass step scaling method itself has, of course, a significant impact on the 
error of the final result, since the ratios are all multiplied in the end. 
Most results, which were obtained by projecting with GEVP eigenvalues, are compatible 
with results obtained by using only the source with the largest gaussian smearing.

We also observed that small momenta injected by twisted-boundary conditions provide a stable extrapolation to the point of zero recoil. 

Finally, as can be seen in Fig.\,\ref{fig_ratiofit}, the data are not sensitive to the change in the heavy mass $M_{h_is}$.

\section*{Acknowledgment}
This project is supported by Agence Nationale de la Recherche under the contract ANR-17-CE31-0019 (B.B. and J.N.).
This work was granted access to the HPC resources of CINES and IDRIS (2018-A0030506808, 2019-A0050506808, 2020-A0070506808 and 2020-A0080502271) by GENCI.
We also gratefully acknowledge the computing time granted on SuperMUC-NG
(project ID {pn72gi}) by the Leibniz Supercomputing Centre of the
Bavarian Academy of Sciences and Humanities at Garching near Munich and
thank its staff for their support. The authors are grateful to the colleagues of the CLS effort for having provided the gauge field ensembles used in the present work.
This work is supported by the Deutsche Forschungsgemeinschaft (DFG) through the Research Training Group “GRK
2149: Strong and Weak Interactions – from Hadrons to Dark Matter” (J.N. and J.H.).

\bibliographystyle{JHEP}
\bibliography{paper}

\end{document}